\documentstyle[prb,aps,multicol,epsf]{revtex}
\tighten
\narrowtext
\newcommand \be {\begin{equation}}
\newcommand \ee {\end{equation}}
\newcommand \bea {\begin{eqnarray}}
\newcommand \eea {\end{eqnarray}}
\begin{document}
\draft

\title{Magnetotransport in lateral superlattices with small-angle
impurity scattering: Low-field magnetoresistance}
\author{A.~D.~Mirlin$^{a,b,*}$, E.~Tsitsishvili$^{b,c}$, 
and P.~W\"olfle$^{a,b}$}
\address{
$^a$ Institut f\"ur Nanotechnologie, Forschungszentrum Karlsruhe,
D-76021 Karlsruhe, Germany\\
$^b$ Institut f\"ur Theorie der Kondensierten Materie,
Universit\"at Karlsruhe, D-76128 Karlsruhe, Germany\\
$^c$ Institute of Cybernetics, Euli 5, 380086 Tbilisi, Georgia}
\date{\today}
\maketitle
\begin{abstract}
An analytical study of the low-field magnetoresistance of a
two-dimensional electron gas subject to a weak periodic modulation is
presented. We assume small-angle impurity scattering
characteristic for high-mobility semiconductor heterostructures. It is
shown that the condition for existence of the strong low-field
magnetoresistance induced by so-called channeled orbits is
$\eta^{3/2}ql\gg 1$, where $\eta$ and $q$ are the strength
and  the wave vector of the modulation, and $l$ is the transport mean
free path. Under this condition, the magnetoresistance scales as
$\eta^{7/2}$.

\end{abstract}

\begin{multicols}{2}

\section{Introduction}
\label{s1}

The effect of a periodic modulation (lateral superlattice) on the
transport properties of a two-dimensional (2D) electron gas has been
the subject of intensive research during the last decade. This
interest was triggered by an experiment of Weiss {\it et al.}
\cite{weiss} who discovered that a weak one-dimensional modulation
(grating) with a wave vector ${\bf q}\parallel {\bf e}_x$ induces
strong commensurability oscillations of the magnetoresistivity
$\rho_{xx}(B)$, while showing almost no effect on $\rho_{yy}(B)$ and
$\rho_{xy}(B)$. The oscillation minima have been found to satisfy the
condition $2R_c/a=n-1/4$ with integer $n$, where $R_c$ is the
cyclotron radius and $a=2\pi/q$ the grating wave length. While these
findings can be explained in terms  of the quantum-mechanical band
structure induced by the modulation \cite{theory-qm},  the phenomenon
is, in fact, of quasiclassical nature, as was demonstrated by
Beenakker \cite{beenakker}. He showed that the geometric resonance of
the cyclotron motion in the grating induces an extra contribution to
the drift velocity of the guiding center, whose root-mean-square
amplitude oscillates as $|\cos(qR_c-\pi/4)|$. These arguments were
corroborated by an analytical solution of the Boltzmann equation
within an expansion in the relative strength $\eta$ of the modulation,
the result for the modulation-induced oscillatory magnetoresistivity
$\Delta\rho_{xx}$ being of the order  $\eta^2$. However, although the
theoretical results \cite{theory-qm,beenakker}  accounted for the
above experimental features,  they disagreed strongly with the
experiment as far as the damping of the oscillations with decreasing
magnetic field is concerned. The reason for this was an oversimplified
treatment of the impurity scattering: while the theory of
\cite{theory-qm,beenakker} assumed isotropic scattering, in
experimentally relevant high-mobility samples the random potential is
very smooth, so that the scattering is of small-angle character, with
the total relaxation rate $\tau_s^{-1}$ much exceeding the momentum
relaxation rate $\tau^{-1}$. This gap in the theory was filled in by
our previous paper \cite{mw98} where the effect of small-angle
scattering on the Weiss oscillations was studied analytically. It was
found that the small-angle scattering changes the dependence of the
oscillation amplitude on the magnetic field $B$ completely, leading to
a much stronger damping of oscillations with decreasing $B$, in very
good quantitative agreement with experimental data.

This paper continues our investigation of the effect of small-angle
scattering on the transport in laterally modulated structures. We will
address the issue of the low-field magnetoresistivity, which has been
left aside in our earlier paper \cite{mw98}.  In combination,
Ref.~\onlinecite{mw98} and the present paper provide a complete
quasiclassical theory of magnetoresistivity of a 2D electron gas
subject to a weak one-dimensional modulation and a smooth random
potential.

A distinct low-field magnetoresistivity was observed, along with the
commensurability oscillations, in the original experiment
\cite{weiss}, as well as in numerous later experiments on the
transport in a lateral superlattice. Specifically,  in low magnetic
fields $B$ a positive magnetoresistivity was found, followed by a
maximum in $\rho_{xx}(B)$. For not too strong modulation, the relevant
magnetic fields are much weaker than those where the Weiss
oscillations are observed, so that the two effects can be easily
separated. Soon after the first experimental observation it was
understood \cite{beton90} that the low-field magnetoresistivity is
related to the existence of open (channeled) orbits in the magnetic
fields $B<B_c=(\eta c/2e)qmv_F$. It is worth mentioning that this
effect, which is not found within the  $\eta$-expansion used in
Refs.~\onlinecite{beenakker,mw98}, has its counterpart in the context
of the sound absorption in metals in the presence of a magnetic
field. There, the trapping of electrons in channeled  orbits by a
sound wave leads to non-linearity of the acoustic response of an
electron gas, as was observed experimentally \cite{fil} and analyzed
theoretically \cite{galperin}. Though the work by Beton {\it et al.}
\cite{beton90} (see also a more recent paper \cite{menne98})
explained qualitatively the low-field magnetoresistivity as an effect
of channeled orbits, a quantitative analytical description of the
problem requires that the nature of disorder be taken into
account. This is done below in this paper, where we demonstrate that
the small-angle character of scattering leads to a parametrically
different magnitude of this effect. We will also study how the
contribution of drifting orbits to $\Delta\rho_{xx}$ is modified in
the low-magnetic-field range, $B\sim B_c$.

\section{Low-field magnetotransport in smooth disorder}
\label{s2}

\subsection{Generalities}
\label{s2a}

\begin{figure}
\narrowtext
{\epsfxsize=7cm\centerline{\epsfbox{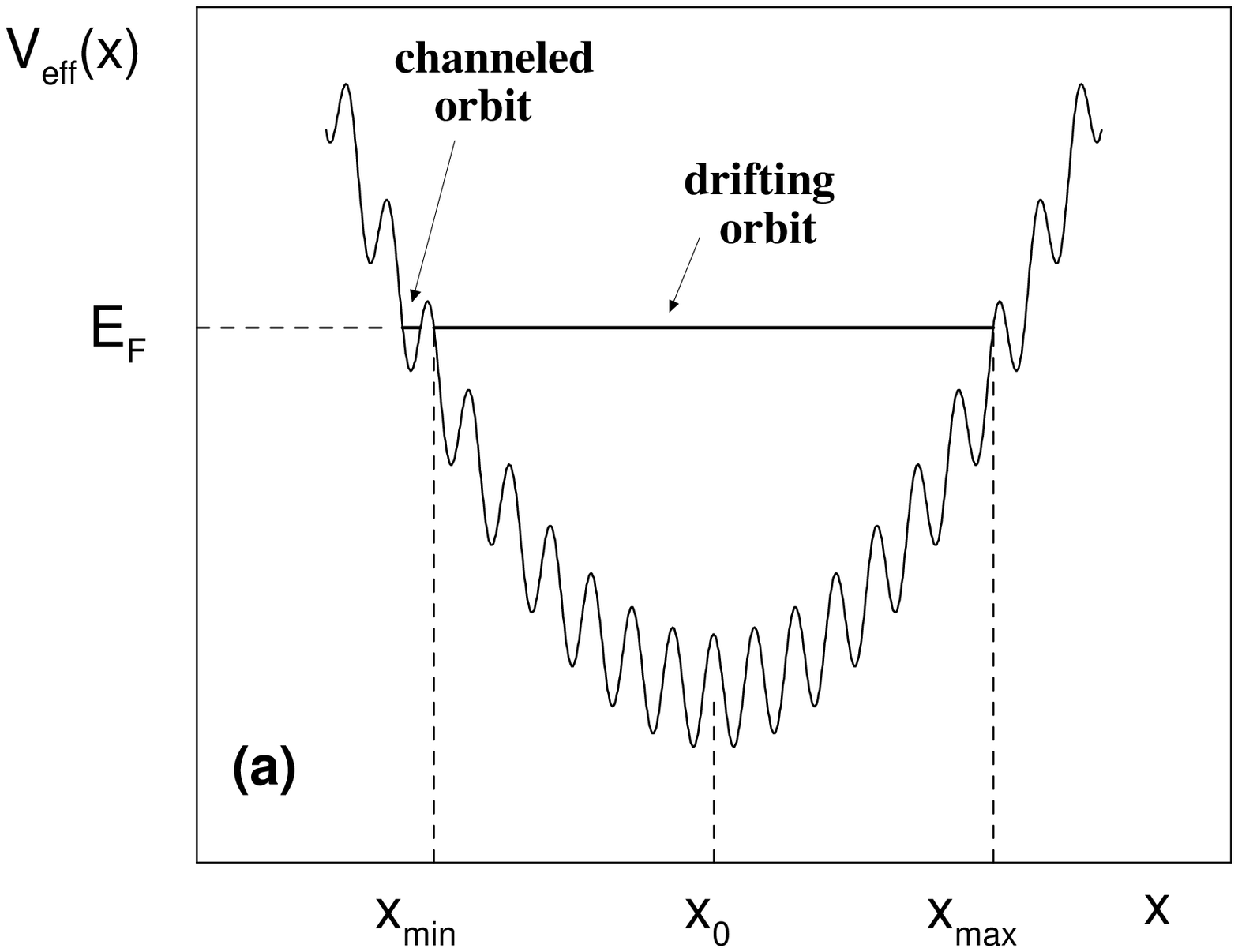}}}
\vskip0.3cm
{\epsfxsize=7cm\centerline{\epsfbox{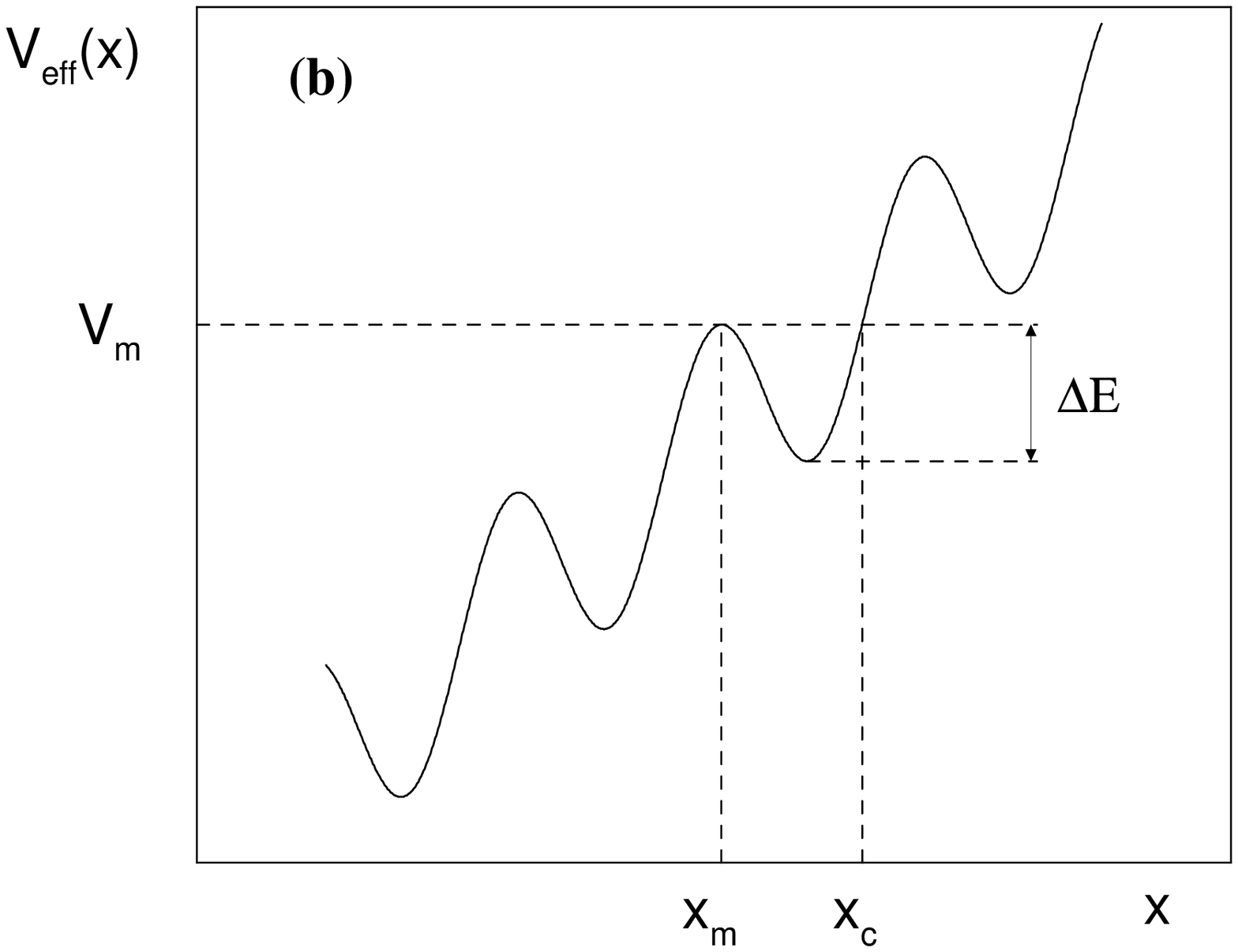}}}
\vskip0.3cm
\caption{(a) Effective potential $V_{\rm eff}(x)$, Eq.~(\ref{e2.8}); (b)
effective potential linearized near a turning point,
Eq.~(\ref{e2.14a}).}
\label{fig1}
\end{figure}

We consider a weak periodic potential
\be
\label{e2.1}
V(x)=\eta E_F \cos qx\ ,\qquad \eta\ll 1
\ee
acting on a 2D electron system (Fermi energy $E_F$) at temperature
$T=0$. In the presence of a perpendicular magnetic field $B$, the
electron at the Fermi level will perform cyclotron motion (frequency
$\omega_c=eB/mc$, cyclotron radius $R_c=v_F/\omega_c$, where $v_F$ is
the Fermi velocity). The periodic potential causes a slow drift motion
in the $y$-direction with velocity $v_d$. The classical equations of
motion are (the electron charge is $-e$) 
\bea
m\ddot{x}&=&-{e\over c}\dot{y}B-{dV\over dx}\ , \label{e2.2}\\
m\ddot{y}&=&{e\over c}\dot{x}B\ . \label{e2.3}
\eea
Averaging (\ref{e2.2}) over one cyclotron revolution one finds the
drift velocity
\be
\label{e2.5}
v_d=-{c\over eB}{\omega_c\over\pi}\int_{x_{\rm min}}^{x_{\rm max}}
{dx\over\dot{x}}{dV\over dx}\ ,
\ee
where $x_{\rm min}$, $x_{\rm max}$ are the classical turning points of
the periodic motion. Integration of (\ref{e2.3}) yields
\be
\label{e2.6}
v_y\equiv \dot{y}=\omega_c(x-x_0)\ ,
\ee
where $x_0$ is the point on the trajectory where $\dot{y}=0$ (i.e. the
guiding center coordinate). The $x$-component of velocity is now
easily found from energy conservation,
\be
v_x\equiv \dot{x}=\left[{2\over m}(E-V_{\rm eff}(x))\right]^{1/2}\ ,
\label{e2.7}
\ee
where the effective potential of the motion in $x$-direction is given
by 
\be
\label{e2.8}
V_{\rm eff}(x)=V(x)+{m\over 2}\omega_c^2(x-x_0)^2 \ .
\ee
For weak magnetic fields, $qR_c\ll 1$, the effective potential is
given by a parabola centered at $x_0$, with rapid periodic
oscillations superposed (Fig.~\ref{fig1}a). 
Intersection of this curve with a line of
constant energy yields the turning points of the classical
motion. There is always a central allowed region $x_{\rm min}<x<x_{\rm
max}$, for which the motion consists of a drift of complete cyclotron
orbits. For sufficiently strong periodic potential, there exists also
additional classically allowed region, to the left (right) of the
turning point $x_{\rm min}$ ($x_{\rm max}$). These are so-called
channeled orbits, winding in a snake-like fashion along $y$. Using the
fact that $x-x_0\simeq R_c$ in the region of channeled orbits, one
finds the condition for the existence of channeled orbits $B<B_c$,
where 
\be
\label{bc}
{e\over c}B_c={\eta mv_Fq\over 2}\ .
\ee 

We will assume that $\omega_c\tau\gg 1$ in the relevant range of
magnetic fields, which implies for $B\sim B_c$ that $\eta\gg
2/ql$, where $l=v_F\tau$ is the transport mean free path. Since
$ql\sim 300 \div 1000$ in a typical experiment with a 
high mobility sample, this condition is fulfilled even for a very weak 
($\eta$ of order of few percent) modulation.

\subsection{Contribution of drifting orbits}
\label{s2b}

We will first discuss the contribution of the drifting cyclotron
orbits to the transport. Substituting (\ref{e2.7}), (\ref{e2.8}) into
(\ref{e2.5}), we find 
\be
\label{e2.9}
v_d={\eta\over 2\pi}v_F\int_{u_{\rm min}}^{u_{\rm max}}du \sin u
\left[1-\eta\cos u-{(u-u_0)^2\over (qR_c)^2}\right]^{-1/2},
\ee
where $u=qx$, $u_0=qx_0$, etc. In the limit $\eta qR_c\ll 1$ (which is
equivalent to 
$B\gg B_c$) the term $\eta\cos u$ in (\ref{e2.9}) may be neglected. The
integral may then be done, yielding
\be
\label{e2.10}
v_d={1\over 2}\eta v_F qR_c J_0(qR_c) \sin u_0\ ,
\ee
where $J_0(x)$ is the Bessel function. The drift velocity is
oscillating with $u_0$, i.e. the initial conditions. In order to
determine the modulation-induced correction to the diffusion
coefficient, one has to evaluate the average of $v_d^2$,
\bea
\label{e2.11}
\langle v_d^2\rangle &\equiv& {1\over 2\pi}\int_0^{2\pi}du_0 v_d^2
={1\over 8}\eta^2v_F^2(qR_c)^2J_0^2(qR_c) \nonumber \\
&\simeq& {1\over 4\pi}\eta^2v_F^2 qR_c \cos^2(qR_c-\pi/4)
\eea
(in the second line we used the condition $qR_c\gg 1$). The result
(\ref{e2.11}) obtained for the first time by Beenakker
\cite{beenakker} induces an oscillating (with $qR_c$) correction 
$\delta \rho_{xx}$ to the resistivity 
along the modulation wave vector (Weiss
oscillations). The amplitude of Weiss oscillations for the case of
white-noise disorder was calculated in \cite{beenakker} and for the
(experimentally relevant) case of a smooth random potential in
\cite{mw98}. Here we are interested in the behavior of 
$\Delta \rho_{xx}$ in the region of sufficiently low magnetic fields,
where the oscillations are exponentially damped. This means that the
position $u_0$ of the guiding center changes due to impurity
scattering by an amount $\delta u_0\gg q^{-1}$ within one cyclotron
revolution. Taking into account that (in view of the oscillatory
factor $\sin u$) the main contribution to the
integral (\ref{e2.9}) comes from the regions close to the turning
points $u_{\rm min}$, $u_{\rm max}$, one easily realizes that an
average squared drift velocity in one
cyclotron revolution is then given by non-oscillatory part of
Eq.~(\ref{e2.11}), 
\be
\label{e2.12}
\langle v_d^2\rangle =  {1\over 8\pi}\eta^2v_F^2 qR_c \ .
\ee
The diffusive process in $y$-direction with the velocity squared
$\langle v_d^2\rangle$ and the time step $\Delta t =2\pi/\omega_c$
induces a correction to the diffusion constant 
\be
\label{e2.13}
\Delta D_{yy}=\langle v_d^2\rangle {\pi\over\omega_c}
\ee
and thus the resistivity correction 
\be
\label{e2.14}
{\Delta\rho_{xx}\over\rho_0}=
{\Delta D_{yy}\over v_F^2\tau/2(\omega_c\tau)^2}= {1\over 4}\eta^2ql\ ,
\ee
where $\rho_0$ is the Drude resistivity in the absence of modulation.
Eq.~(\ref{e2.14}) reproduces the result of
Refs.~\onlinecite{beenakker,mw98} 
for the saturation value of ${\Delta\rho_{xx}/\rho_0}$ at low magnetic
fields. 

Having demonstrated how the result (\ref{e2.14}) of $\eta$-expansion
of the Boltzmann equation is reproduced within the present approach
for $B\gg B_c$, 
we are prepared to turn to the question of our main interest here,
{\it i.e.} to 
the range of lower magnetic fields, $B\sim B_c$. Since the main
contribution to the drift velocity, Eq.~(\ref{e2.9}), comes from the
vicinity of the turning points, one can linearize the second
(parabolic) term in the effective potential (\ref{e2.8}) near the
points $x=x_{0\pm}\equiv x_0\pm R_c$,
\bea
\label{e2.14a}
V_{\rm eff}(x)-E_F &=& V(x)\pm mv_F\omega_c(x-x_{0\pm}) \nonumber\\
                   &=& \eta E_F[\cos u\pm\beta(u-u_{0\pm})]\ .
\eea
We have introduced here the parameter
\be
\label{e2.17}
\beta = {B\over B_c}\ .
\ee
The contribution of the lower limit takes then
the form
\be
\label{e2.15}
v_d^{\rm min} \simeq  {\eta v_F\over 2\pi}
\left({qR_c\over 2}\right)^{1/2} F(u_0,\beta)\ ,
\ee
where
\be
\label{e2.16}
F(u_0,\beta)=\int_{u_{\rm min}}^\infty du\,\sin u
\left(u-u_{0-}-{1\over\beta}\cos u\right)^{-1/2}\ .
\ee
The upper limit contribution is found to be 
$v_d^{\rm max}=-v_d^{\rm min}|_{u_0\to -u_0}$. The average value of
the total drift velocity $v_d=v_d^{\rm max}+v_d^{\rm min}$ is
therefore equal zero, $\langle v_d\rangle=0$, where, as before the
averaging goes over the position of the guiding center, 
$$\langle\ldots\rangle ={1\over 2\pi}\int_0^{2\pi}du_0\ldots\ .$$
The average of the square of $v_d$ is obtained as
$$\langle v_d^2\rangle = 2[\langle(v_d^{\rm min})^2\rangle - 
\langle v_d^{\rm min} \rangle^2]\ , $$
where we used $\langle(v_d^{\rm min})^2\rangle=
\langle(v_d^{\rm max})^2\rangle$ and 
$\langle v_d^{\rm min}v_d^{\rm max} \rangle \simeq 
\langle v_d^{\rm min}\rangle \langle v_d^{\rm max} \rangle$. The
latter relation is ensured by the impurity scattering leading to a
large shift $\delta u_0\gg q^{-1}$ within one cyclotron revolution, as
has been discussed above.

\begin{figure}
\narrowtext
{\epsfxsize=7cm\centerline{\epsfbox{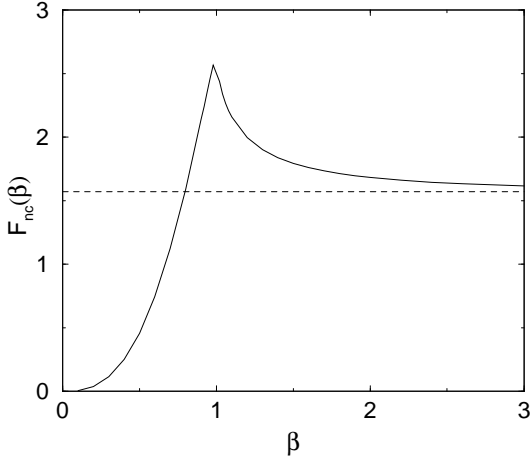}}}
\caption{Function $F_{\rm nc}(\beta)$ characterizing the magnetic
field dependence of the contribution of non-channeled orbits to
resistivity. The dashed line indicates the asymptotic value 
$F_{\rm nc}(\beta\gg 1)=\pi/2$ corresponding to the result
(\ref{e2.14}) of the $\eta$-expansion.}
\label{fig2}
\end{figure}

Using (\ref{e2.13}), (\ref{e2.14}), we obtain thus the following
contribution of the drifting orbits to the resistivity correction
\be
\label{e2.18}
{\Delta\rho_{xx}^{\rm nc}\over\rho_0}=2\pi\omega_c\tau{\langle
v_d^2\rangle\over v_F^2}={\eta^2\over 2\pi} ql F_{\rm nc}(B/B_c)\ ,
\ee
where the dimensionless function $F_{\rm nc}(\beta)$ is defined by
\be
\label{e2.19}
F_{\rm nc}(\beta)=\langle F^2(u_0,\beta)\rangle-
\langle F(u_0,\beta)\rangle^2\ .
\ee
The function $F_{\rm nc}(\beta)$ is shown in Fig.~\ref{fig2} . In the limit 
$\beta\to \infty$ one easily finds $F_{\rm nc}(\beta)\to \pi/2$, 
reproducing Eq.~(\ref{e2.14}). 

\subsection{Contribution of channeled orbits}
\label{s2c}

\begin{figure}
\narrowtext
{\epsfxsize=7cm\centerline{\epsfbox{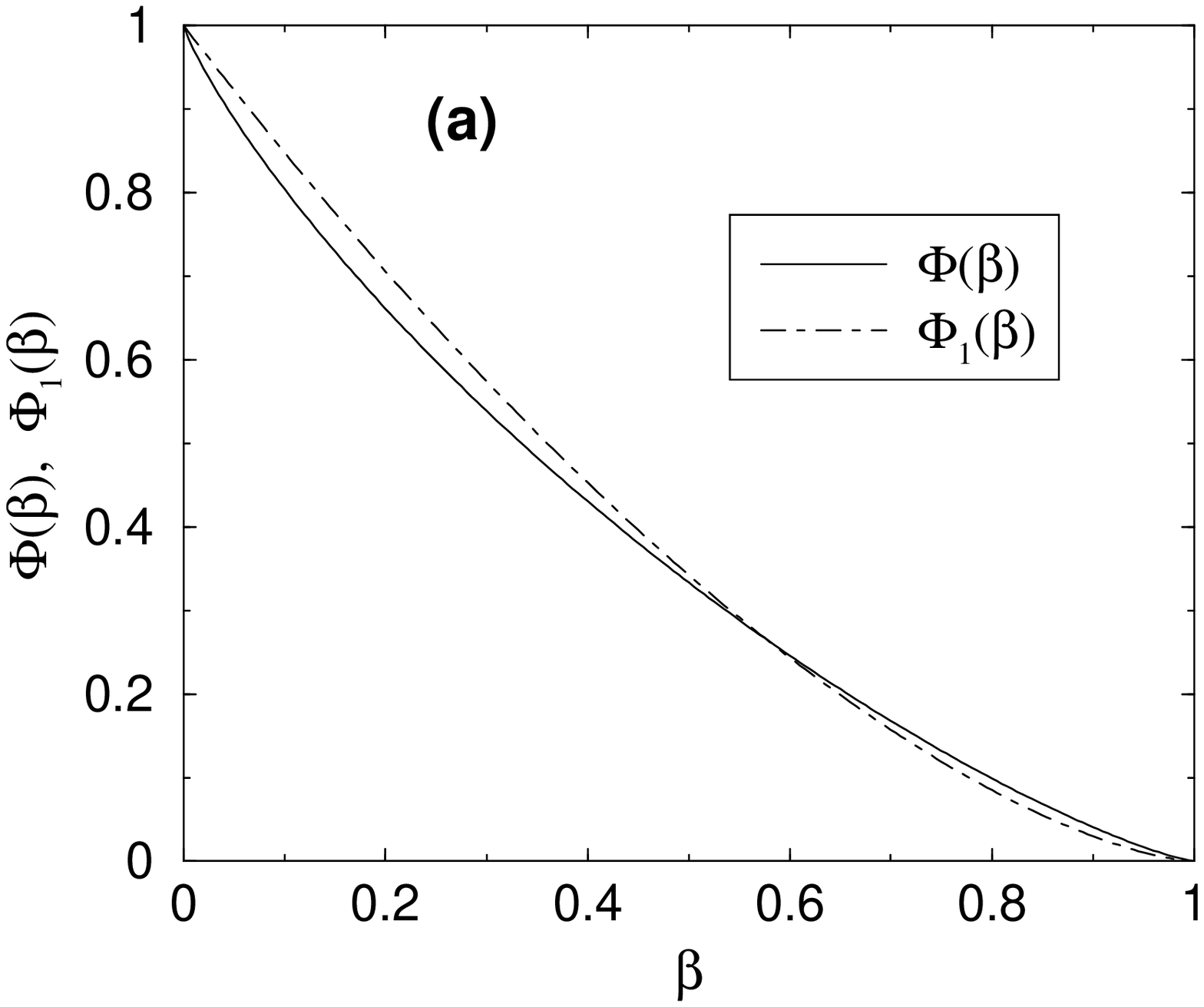}}}
{\epsfxsize=7cm\centerline{\epsfbox{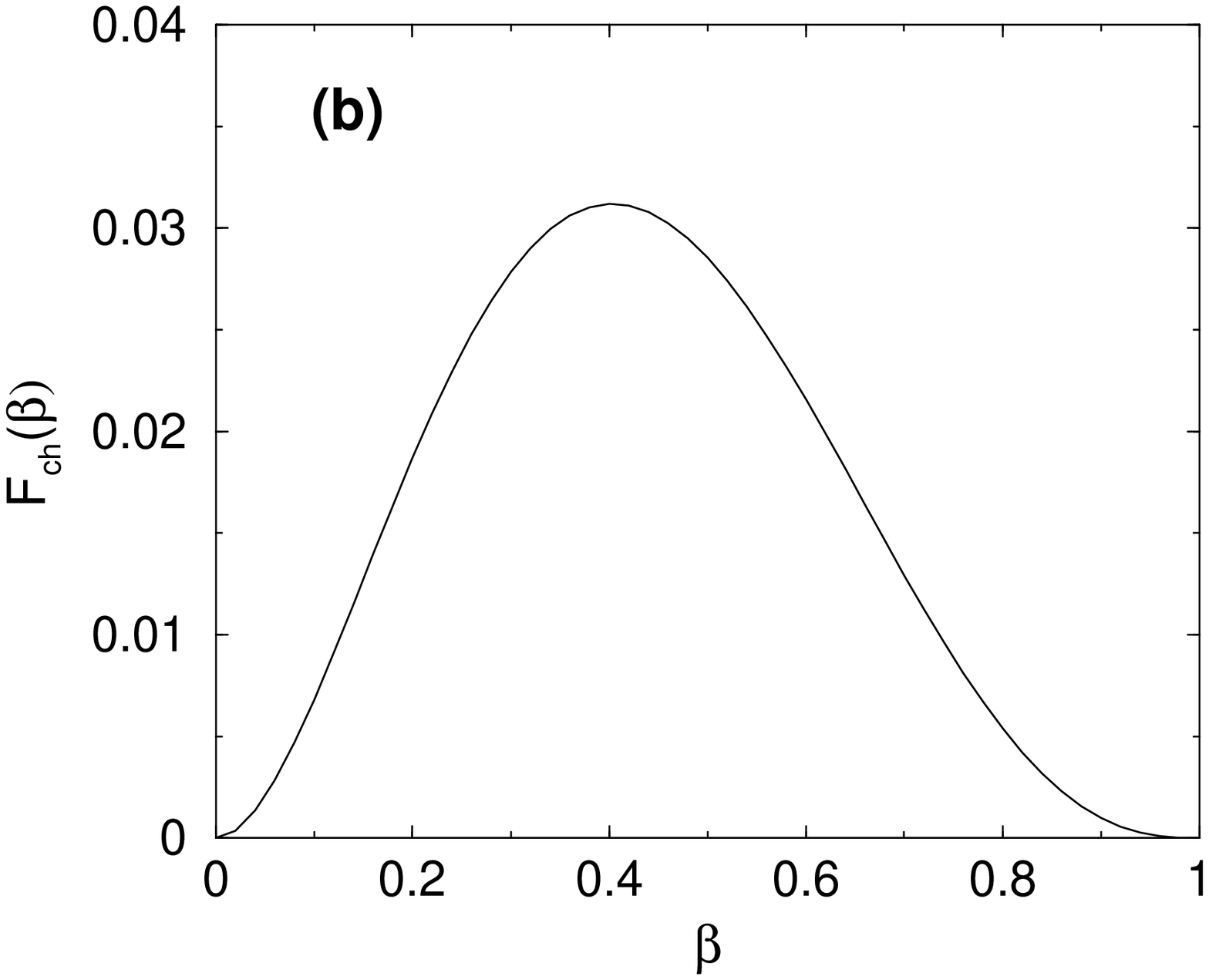}}}
\caption{a) Functions $\Phi(\beta)$ and $\Phi_1(\beta)$
characterizing the fraction of open orbits and the life time in a
channel; b) function $F_{\rm ch}(\beta)$ describing magnetic field
dependence of the contribution of channeled orbits to the resistivity,
Eq.~(\ref{e2.36}).} 
\label{fig3}
\end{figure}

We now turn to the contribution of channeled orbits. A particle will
spend an average time $\tau_{\rm ch}$ in a channeled orbit and during this
time will propagate with velocity $v_F$. If the fraction of particles
in channeled orbits is $P_{\rm ch}(\beta)$, the contribution to the
diffusion coefficient will be
\be
\label{e2.20}
\Delta D_{yy}^c ={1\over 2}v_F^2\tau_{\rm ch}P_{\rm ch}(\beta)\ .
\ee
In order to calculate $P_{\rm ch}(\beta)$ (see
e.g. Refs.~\onlinecite{beton90,menne98}), we consider again the effective 
potential $V_{\rm eff}(x)$ linearized near a turning point,
Eq.~(\ref{e2.14a}). The fraction of the phase space $(x,\phi)$
occupied by the channeled orbits is clearly given by
\be
\label{e2.22}
P_{\rm ch}(\beta) = \int_{x_m}^{x_c}{dx\over a} {2\phi_{\rm max}(x)\over
\pi}\ ,
\ee
where $x_m$ is the position of a maximum of $V_{\rm eff}$, $x_c$ is
the nearest point to $x_m$ 
satisfying $V_{\rm eff}(x_c)=V_{\rm eff}(x_m)$ (see
Fig.~\ref{fig1}b), and $\phi_{\rm 
max}(x)$ is the limiting angle of the velocity vector with the
$y$-axis. In view of $\eta\ll 1$ we have $\phi_{\rm max}\ll 1$, so
that
\be
\label{e2.23}
\phi_{\rm max}(x)=\left[{V_m-V_{\rm eff}(x)\over
mv_F^2/2}\right]^{1/2}\ , \qquad V_m=V_{\rm eff}(x_m)\ .
\ee
Substitution of (\ref{e2.23}) into (\ref{e2.22}) yields
\be
\label{e2.24}
P_{\rm ch}(\beta)= {4\over \pi^2}(2\eta)^{1/2}\Phi(\beta),
\ee
where
\bea
\label{e2.25}
\Phi(\beta) & = & {1\over 4\sqrt{2}}
\int_{\arcsin \beta}^{\arcsin \beta+2\pi} du \nonumber \\
& \times & {\rm Re} [\sqrt{1-\beta^2}+\beta\arcsin\beta-
\cos u -\beta u]^{1/2}\ .
\eea
The function $\Phi(\beta)$ satisfies $\Phi(0)=1$ and $\Phi(1)=0$ and
is shown in Fig.~\ref{fig3}a.

The life time of a particle in a channeled orbit, $\tau_{\rm ch}$, may
be obtained by considering the diffusion in the space $(\dot{x},x)$ in
the potential well formed by the potential $V_{\rm eff}(x)$. Particles
diffusing beyond the borderline marked by the maximum of 
$V_{\rm eff}(x)$ escape the confining well. The lifetime of particles
may be defined from the time dependence of the total number of
particles in the well, $n_w(t)$, as
\be
\label{e2.26}
\tau_{\rm ch}= \int_0^\infty dt\, n_w(t)\ ,
\ee
where it is assumed that at $t=0$ there is one particle in the well,
$n_w(0)=1$, spread uniformly over the volume $\Omega$ of the
corresponding phase space. In
order to calculate $n_w$ we define the phase space density
$f(x,\dot{x},t)$, subject to the initial condition
$f(x,\dot{x},0)=1/\Omega$ and to the boundary condition $f=0$ at the
boundary of the confinement region. 

To determine $f(x,\dot{x},t)$ analytically, we approximate the
potential well by a parabola
\bea
\label{e2.27}
V_{\rm eff}(x)=4 \Delta E
{[x-(x_c+x_m)/2]^2\over (x_c-x_m)^2} + {\rm const}\ , && \\
 x_m<x<x_c\ , && \nonumber 
\eea
where $\Delta E$ is the depth of the well. Introducing dimensionless
variables
\be
\label{e2.28}
X={2[x-(x_c+x_m)/2]\over (x_c-x_m)}\ ,\ \ \  
Y=\left({m\over 2\Delta E}\right)^{1/2}\dot{x}\ ,
\ee
and $\epsilon=E/\Delta E$, the energy takes the isotropic form
\be
\label{e2.29}
\epsilon=X^2+Y^2\ ,\qquad \epsilon\le 1\ .
\ee
The scaled phase space trajectories are thus circles. Let us now
consider the effect of impurity scattering. The small-angle scattering
by the smooth random potential induces a diffusive motion in the space
of the velocity angle, 
$\langle (\delta\phi)^2\rangle=2D_\phi\delta t$, with the diffusion
coefficient $D_\phi=1/\tau$, where $\tau$ is the transport time. 
For channeled orbits we have $\dot{x}\simeq\phi v_F$, implying a
diffusion in $\dot{x}$ with diffusion coefficient
$D_{\dot{x}}=v_F^2/\tau$. Provided the motion of the
particles in the potential well is rapid as compared to the escape
process (requiring the condition $\eta^{3/2}ql\gg 1$, see below), the
effect of the anisotropic diffusion (along $Y$) is equivalent to
isotropic diffusion with diffusion coefficient
\be
\label{e2.30}
D={1\over 2}D_Y={m\over 4\Delta E}{v_F^2\over\tau}\ .
\ee
We are thus left with the diffusion equation in a circle,
\be
\label{e2.31}
D\nabla^2 n(R,t) - {\partial n\over \partial t} = 0\ ,\qquad
0\le  R\equiv (X^2+Y^2)^{1/2} \le 1\ ,
\ee
supplied with the boundary condition $n(1,t)=0$ and the initial
condition $n(R,0)=1/\pi$. The solution is easily found by expanding
$n(R,t)$ in eigenfunctions $J_0(\kappa_n R)$ of the Laplace operator,
\be
\label{e2.32}
n(R,t)={2\over\pi}\sum_{n=1}^\infty {1\over \kappa_n J_1(\kappa_n)}
J_0(\kappa_n R)e^{-D\kappa_n^2t}\ ,
\ee
where $\kappa_n$ are zeros of the Bessel function $J_0(x)$. 
We find thus for the integrated staying probability
$n_w(t)=2\pi\int_0^1 dR R\, n(R,t)$
\be
\label{e2.33}
n_w(t)=\sum_n{4\over\kappa_n^2}e^{-D\kappa_n^2t}\ .
\ee
The lifetime of a particle in the channel is now found from 
(\ref{e2.26}) as
\be
\label{e2.34}
\tau_{\rm ch}={4\over D}\sum_n{1\over\kappa_n^4}={1\over 8D} = 
{1\over 2}\eta \tau\Phi_1(\beta)\ ,
\ee
where
\be
\label{e2.35}
\Phi_1(\beta)=\sqrt{1-\beta^2} + \beta\arcsin\beta 
-{\pi\over 2}\beta\ .
\ee
The dimensionless function $\Phi_1(\beta)$ decreasing monotonically
from 1 at $\beta=0$ to 0 at $\beta=1$ is shown in Fig.~\ref{fig3}a.

Collecting the results for $P_{\rm ch}(B)$ and $\tau_{\rm ch}$ we find
the correction to the resistivity due to transport in channeled
trajectories as
\bea
\label{e2.36}
{\Delta\rho_{xx}^{\rm ch}\over\rho_0} & = & 
2(\omega_c\tau)^2 P_{\rm ch}(\beta) \tau_{\rm ch}/\tau \nonumber \\
& = & {\sqrt{2}\over \pi^2}\eta^{7\over 2}(ql)^2 F_{\rm ch}(\beta)\ ,
\eea
where
\be
\label{e2.37}
F_{\rm ch}(\beta)=\beta^2\Phi_1(\beta)\Phi(\beta)\ .
\ee
Therefore, the magnitude of the low-field magnetoresistance (i.e. the
value of $\Delta\rho_{xx}^{\rm ch}/\rho_0$ at maximum) scales as
$\eta^{7/2}$ with the modulation amplitude. 
The function $F_{\rm ch}(\beta)$ approaches zero for $\beta\to 0$ and
$\beta\to 1$ and has a maximum of $\sim 0.03$ at $\beta\simeq 0.4$ (see
Fig.~\ref{fig3}b). 

\subsection{Resistivity correction at zero magnetic field}
\label{s2d}

Formulas (\ref{e2.18}), (\ref{e2.36}) give a zero resistivity
correction in the limit $\beta\to 0$. However, these formulas are not
valid for too small $\beta$, and in particular, at $B=0$. This is
clear already from the fact that we used the condition
$\omega_c\tau\gg 1$ in the course of their derivation, which implies
$\beta\gg 1/\eta ql$. (In fact, the lower limit for $\beta$ is even
more restrictive, see Eq.~(\ref{e2.39}) below). Therefore, the zero
magnetic 
field case requires a separate treatment. Our approach, based on
consideration of a disorder-induced diffusion in the space of electron
orbits, can be, however, generalized to the case $B=0$ as well. The
calculation presented in Appendix yields the following result:
\be
\label{e2.38}
{\Delta\rho_{xx}(B=0)\over\rho_0} = C\eta^{1/2}\ ,
\ee
where $C\simeq 1.24$ is a numerical coefficient. How does
Eq.~(\ref{e2.38}) match the finite-$B$ result (\ref{e2.36})? In our 
derivation in Sec.~\ref{s2b}, \ref{s2c} we assumed that the cyclotron
motion away from the domain of the phase space occupied by the channeled
orbits is faster than the escape process from a channel. Since
a characteristic angle  $\phi_{\rm max}$ for the channeled orbit is
$\phi_{\rm max}\sim\eta^{1/2}$, we get the condition $\tau_{\rm
ch}\gg\eta^{1/2}/\omega_c$, which can be rewritten as
\be
\label{e2.39}
\beta\gg {1\over \eta^{3/2}ql}\ .
\ee
Note that the r.h.s. of (\ref{e2.39}) is assumed to be much smaller
than unity, see Sec.~\ref{s2e}. In the opposite case, 
$\beta\ll 1/\eta^{3/2}ql$, the zero-field result (\ref{e2.38}) should
be valid. Indeed, comparing (\ref{e2.36}) and (\ref{e2.38}), we see
that they match at $\beta\sim 1/\eta^{3/2}ql$.

\subsection{Conditions of applicability}
\label{s2e}

Let us discuss conditions of validity of our consideration. First of
all, the picture of the diffusion in the velocity space used above is
justified if the particle undergoes many scattering events within the
time $\tau_{\rm ch}\sim\eta\tau$, which implies 
\be
\label{e2.40}
\eta\gg \tau_s/\tau.
\ee
Assuming a value $\tau/\tau_s\sim 50$ characteristic
for high mobility samples, this inequality is reasonably fulfilled for
modulation strengths $\eta\gtrsim 5\%$. In the opposite case 
$\eta\ll \tau_s/\tau$ (i.e for very weak modulations) one should
replace $\tau_{\rm ch}$ by $\tau_s$ in Eq.~(\ref{e2.36}), yielding
\be
\label{e2.41}
{\Delta\rho_{xx}^{\rm ch}\over\rho_0}=
{2^{3/2}\over \pi^2}\eta^{5\over 2}q^2ll_s \beta^2\Phi(\beta)\ ,
\ee
where $l_s=v_F\tau_s$, so that the $\eta^{7/2}$ behavior changes into
the $\eta^{5/2}$ one. As to the zero-$B$
formula (\ref{e2.38}), we expect that it transforms at
$\eta\ll\tau_s/\tau$ into  
$\Delta\rho/\rho_0\sim \eta^{3/2}\tau/\tau_s$, matching the result 
$\Delta\rho/\rho_0\sim \eta^{3/2}$ obtained \cite{menne98} in the
limit of isotropic scattering ($\tau_s=\tau$). 

Another essential assumption was that the oscillatory motion within
the potential well takes place on a time scale much shorter than the
escape time $\tau_{\rm ch}$. In other words, we assume that the
particle performs many oscillations in the channel (with the frequency
$\omega_{\rm ch}\sim\eta^{1/2}v_Fq$) before it escapes into a
conventional cyclotron orbit. The condition for this is
$\omega_{\rm ch}\tau_{\rm ch}\gg 1$, which means that
\be
\label{e2.42}
\eta^{3/2}ql\gg 1. 
\ee
For experimentally relevant values of $ql$ this
implies $\eta\gg 1\div2\%$, which is fulfilled in the majority of
experiments. In the opposite case, $\eta^{3/2}ql\ll 1$, disorder
scattering dominates over modulation-induced effects, so that
a particle escapes from the channel before it ``recognizes'' that it
is trapped there.  In this
limit the non-perturbative in $\eta$ effects related to existence of
channeled orbits become irrelevant, and the magnetoresistivity is
given by the result of the perturbative expansion (i.e. by
Eq.~(\ref{e2.14}) in the region of exponentially damped oscillations),
without any pronounced features around $B=B_c$. 

Note that in the assumed regime (\ref{e2.42}) the condition 
$\omega_{\rm ch}\tau_{\rm ch}\gg 1$ is still violated in a narrow
vicinity of $\beta=1$, namely at $1-\beta\lesssim
(\eta^{3/2}ql)^{-4/7}$, since both $\omega_{\rm ch}$ and 
$\tau_{\rm ch}$ vanish when $\beta\to1$, 
$\tau_{\rm ch}\propto(1-\beta)^{3/2}$ and 
$\omega_{\rm ch}\propto(1-\beta)^{1/4}$. This
implies that channeled orbits get gradually destroyed by disorder as
$\beta$ approaches unity and leads to  a smearing of
singularity in $\Delta\rho_{xx}/\rho_0$ at $\beta=1$  [Eqs.~(\ref{e2.18}),
(\ref{e2.36})] over a narrow interval 
$\delta\beta\sim(\eta^{3/2}ql)^{-4/7}$. 

Finally, we discuss the overall magnitude of the effect. At
$B>B_c$ only the drifting orbits exist, and Eq.~(\ref{e2.18}) predicts
an enhancement of $\Delta\rho_{xx}/\rho_0$ [with respect to its value well
above $B_c$, Eq.~(\ref{e2.14})] by the factor $F_{\rm nc}(\beta)\sim
1$. At $B<B_c$ comparison of Eq.~(\ref{e2.36}) and Eq.~(\ref{e2.18})
shows that the contribution of channeled orbits dominates, yielding an
enhancement of $\Delta\rho_{xx}/\rho_0$ [again with respect to
(\ref{e2.14})] by a parametrically large factor
$(2^{5/2}/\pi^2)F(\beta)\eta^{3/2}ql$. Let us note, however, that in
view of a rather small numerical value of $F(\beta)$ (equal to 0.03
at the maximum), this enhancement factor may be not so big, despite a
large value of the parameter $\eta^{3/2}ql$.

\section{Summary}
\label{s4}

We have presented an analytical study of the low-field
magnetoresistance of a 2D electron gas subject to a weak
one-dimensional modulation. We assumed that the disorder scattering is
of small-angle nature due to the long-range character of the random
potential. This corresponds to the experimental situation in high-mobility
semiconductor heterostructures, where the smoothness of the impurity
potential is controlled by the large spacer separating the doping layer
from the 2D electron gas. 

We have demonstrated that a strong magnetoresistance with a pronounced
maximum at $B\sim 0.4 B_c$ (with $B_c$ defined by Eq.~(\ref{bc}))
exists for sufficiently strong modulation, $\eta^{3/2}ql\gg 1$. (In
fact, due to smallness of a numerical coefficient, a rather large
value of this parameter is required, $\eta^{3/2}ql\gtrsim 50$.) The
amplitude of $\Delta\rho$ scales with the modulation strength $\eta$
in this regime as $\eta^{7/2}$, see Eq.~(\ref{e2.36}). At zero
magnetic field the grating-induced correction to the resistivity is
small and scales as $\eta^{1/2}$, see Eq.~(\ref{e2.38}). At $B\gg B_c$
the magnetoresistivity is described by the earlier theory using the
perturbative expansion in $\eta$ (Ref.~\onlinecite{beenakker} for
white-noise random potential and Ref.~\onlinecite{mw98} for smooth
disorder).

We have presented a detailed discussion of the limits of validity of
the theory and, in particular, its matching with the earlier results
of the $\eta$-expansion \cite{mw98}. Specifically, for a sufficiently
weak modulation, $\eta^{3/2}ql\ll 1$, the result of the
$\eta$-expansion (predicting $\Delta\rho\propto \eta^2$) is valid in
the whole range of magnetic fields including the low-field region
$B\lesssim B_c$, implying disappearance of the strong low-field
magnetoresistance.  Therefore, the $\eta^{7/2}$ behavior of
$\Delta\rho$ at $B<B_c$, Eq.~(\ref{e2.36}), does not imply a true
non-analyticity of $\Delta\rho(\eta)$ at $\eta\to 0$ but rather
restricts the applicability of the $\eta$-expansion at weak $B$ to the
region of sufficiently small modulation amplitudes, 
$\eta^{3/2}ql\ll 1$. 

Finally, to illustrate the magnitude of the effect, let us calculate 
$\Delta\rho/\rho$ for typical experimental parameters. Specifically,
we use the parameters of a recent experiment \cite{albrecht}: electron
density $n=2.84\times 10^{11}\:{\rm cm}^{-2}$, modulation period
$a=120\:{\rm nm}$, transport mean free path $l=19\:\mu{\rm m}$. While
the emphasis in \cite{albrecht} was put on a novel type of quantum
magnetooscillations, a pronounced magnetoresistance was observed in
low magnetic fields, with a maximum at $B_m=0.145\:{\rm T}$. Using
$B_m\simeq 0.4 B_c$, we infer the modulation amplitude $\eta=0.157$,
so that $\eta^{3/2}ql\simeq 62$ is sufficiently large. With these
values of parameters we find from (\ref{e2.36}) the value of the
magnetoresistivity at maximum, 
$(\Delta\rho_{xx}^{\rm ch}/\rho_0)_{\rm max}\simeq 6.8$, in good
agreement with the experimentally observed magnitude of the effect,       
$\rho_{xx}(B_m)/\rho_{xx}(0)\simeq 8$. As concerns the resistivity
correction at zero magnetic field, Eq.~(\ref{e2.38}), we find 
$\Delta\rho_{xx}(B=0)/\rho_0\simeq 0.49$. Therefore, despite
the parametric smallness of this correction at $\eta\ll 1$, its value
in real experiments can be quite appreciable, due to its square-root
dependence on the modulation strength $\eta$. 

\section*{Acknowledgments}

This work was supported by the SFB195 der Deutschen
Forschungsgemeinschaft, the DFG-Schwerpunktprogram
``Quanten-Hall-Systeme'', the German-Israeli Foundation, and the INTAS 
grant 97-1342. 

\appendix
\section{Resistivity correction at $B=0$}
\label{sa1}

In zero magnetic field and in the absence of disorder equations
(\ref{e2.2}), (\ref{e2.3}) trivially decouple. We denote the 
energies corresponding to the motion along $x$ and $y$ as $E_x$ and
$E_y$, respectively ($E_x+E_y=E_F$). The motion along $x$ is unbound
for $E_x>\eta E_F$; in the opposite case,  $E_x<\eta E_F$, the
particle is in a channeled orbit. The trajectories can be conveniently
labeled by an angle $\psi$ such that $E_x=E_F\sin^2\psi$,
$E_y=E_F\cos^2\psi$. Transforming the uniform distribution on the
Fermi surface $n_\phi^{(0)}(x,\phi)=1/2\pi a$ to the variables
$(x,\psi)$ and integrating over $x$, we find the equilibrium
distribution in the $\psi$-space,
\be
\label{ea1.1}
n_\psi^{(0)}(\psi)={1\over 2\pi}r_-(\psi)\ ,
\ee
where
\bea
\label{ea1.2}
&&r_{\pm}(\psi)=\int {d\vartheta\over 2\pi} 
R^{\pm 1}(\vartheta,\psi)\ , \\
&& R(\vartheta,\psi)=\left({\partial\psi\over\partial\phi}\right)_x
= \left(1-{\eta\over\sin^2\psi}\cos\vartheta\right)^{1/2}\ ,
\label{ea1.3}
\eea
and $\vartheta=qx$. Eq.~(\ref{ea1.2}) is valid for non-channeled
orbits, $\sin^2\psi>\eta$ (consideration of which will be sufficient
for us to determine the conductivity, as explained below). 

The next step is the derivation of a kinetic equation characterizing
the relaxation of $n_\psi(\psi)$ toward equilibrium due to
disorder. The starting 
point is the Liouville-Boltzmann equation for $n_\phi(x,\phi)$, 
\be
\label{ea1.4}
\partial_t n_\phi=-\dot{\phi}\partial_\phi n_\phi-
\dot{x}\partial_x n_\phi + {1\over\tau}\partial_\phi^2 n_\phi\ ,
\ee
where the last term describes the diffusion in the velocity space
induced by the small-angle scattering (in the absence of this term
$\psi$ would be an integral of motion, implying $\partial_t
n_\psi=0$). Now we transform
Eq.~(\ref{ea1.4}) from the variables $(x,\phi)$ to $(x,\psi)$ and use
the fact that the oscillations due to the motion in $x$-direction can
be considered as a fast process as compared to the impurity scattering 
[the corresponding condition is specified in Sec.~\ref{s2e}, see
eq.~(\ref{e2.42})]. This allows us to average over $x$, which yields
the following Fokker-Planck equation for $n_\psi(\psi)$,
\be
\label{ea1.5}
\partial_t n_\psi={1\over\tau}\partial_\psi r_+(\psi)\partial_\psi
{n_\psi \over r_-(\psi)}\ .
\ee

In order to calculate the conductivity, we use the
classical Kubo formula,
\bea
\label{ea1.6}
\sigma_{xx} &=& e^2\nu\int_0^\infty dt \int_{-\infty}^\infty
dx\int_0^{2\pi}d\phi\int_0^a{dx'\over a}\int_0^{2\pi}
{d\phi'\over 2\pi} \nonumber \\
& \times & P(x,\phi;x',\phi';t)v(x)\sin\phi\, v(x')\sin\phi'\ ,
\eea
where $v(x)=[2(E_F-V(x))/m]^{1/2}$ is the particle velocity and
$P(x,\phi;x',\phi';t)$ is the propagator in the phase space
$(x,\phi)$ (i.e. the probability density to move from a point
$(x',\phi')$ to a point $(x,\phi)$ in a time $t$). Transforming from
the variables $(x,\phi)$ to $(\tau,\psi)$, where $\tau$ is the time along
the trajectory, we get
\bea
\label{ea1.7}
\sigma_{xx} &=& e^2\nu\int_0^\infty dt\int d\psi
\int_{-\infty}^\infty d\tau {1\over a}\int {d\psi'\over 2\pi} 
\int_0^{T(\psi')} d\tau'\nonumber \\
& \times & v_x(\psi,\tau)v_x(\psi',\tau')v_F\sin\psi'
P(\tau,\psi;\tau',\psi';t)\ ,
\eea
where $T(\psi)={a\over v_F\sin\psi} r_-(\psi)$ is the period of
oscillations in $v_x$ induced by the modulation. Averaging over the
fast variable $\tau$, we reduce Eq.~(\ref{ea1.7})  to the form
\be
\label{ea1.8}
\sigma_{xx} = e^2\nu v_F^2\int_0^\infty dt\int d\psi
\int {d\psi'\over 2\pi} {\sin\psi\sin\psi'\over r_-(\psi)} 
P(\psi,\psi',t)\ .
\ee
The evolution kernel $P(\psi,\psi',t)$ satisfies the differential
equation (\ref{ea1.5}) with the initial condition
$P(\psi,\psi',0)=\delta(\psi -\psi')$. Defining  
$n(\psi)=\int_0^\infty dt\int d\psi' P(\psi,\psi',t)\sin\psi'$ and
$\tilde{n}(\psi)= n(\psi)/ r_-(\psi)$, we transform Eq.~(\ref{ea1.8})
to the following form:
\bea
&& \sigma_{xx}=e^2\nu v_F^2\int {d\psi\over
2\pi}\sin\psi\,\tilde{n}(\psi)\ ,\label{ea1.9}\\
&&{1\over\tau}\partial_\psi
r_+(\psi)\partial_\psi\tilde{n}(\psi)=-\sin\psi\ . 
\label{ea1.10}
\eea
The range of variation of the variable $\psi$ in (\ref{ea1.9}),
(\ref{ea1.10}) is restricted by the condition of non-channeled motion,
$|\sin\psi|\ge \eta^{1/2}$. As soon as the particle comes into the
region of channeled orbits, its velocity $v_x$ starts to oscillate
rapidly around zero, so that the contribution of such trajectories to
the conductivity can be neglected (the corresponding parameter is
given by Eq.~(\ref{e2.42})). Therefore, Eq.~(\ref{ea1.10}) should be
supplemented by the condition $\tilde{n}=0$ at the boundary of the
non-channeled region. In other words, the integral 
$\int (d\psi/ 2\pi)$ in (\ref{ea1.9}) is understood as 
$\int_{\eta^{1/2}}^{\pi-\eta^{1/2}}(d\psi/\pi)$ (we used $\eta\ll 1$),
and the boundary condition to Eq.~(\ref{ea1.10}) reads
$\tilde{n}(\psi=\eta^{1/2},\:\pi-\eta^{1/2})=0$. 

The function $r_+(\psi)$ is equal to 
$r_+(\psi)\simeq 1-{1\over 16}\eta/\sin^2\psi$ 
for $\sin^2\psi\gg \eta$, reaching the value $2\sqrt{2}/\pi$ at
the boundary, $\sin^2\psi=\eta$. In the limit $\eta\to 0$ we have
$r_+=1$ and $\tilde{n}(\psi)=\tau\sin\psi$, yielding the Drude
conductivity $\sigma_{xx}=e^2\nu v_F^2\tau/2\equiv\sigma_0$. We now
want to calculate 
the leading correction. It will be shown below to be of order
$\eta^{1/2}$, so that we will neglect all contributions of higher
orders. 

Let us denote by $\hat{D}$ the differential operator entering
Eq.~(\ref{ea1.10}), $\hat{D}=-\partial_\psi r_+(\psi)\partial_\psi$
on the interval $[\eta^{1/2},\pi-\eta^{1/2}]$ with zero boundary
conditions, so that Eq.~(\ref{ea1.9}) can be rewritten as
\be
\label{ea1.11}
\sigma_{xx}=e^2\nu v_F^2\tau\int_{\eta^{1/2}}^{\pi-\eta^{1/2}}
{d\psi\over\pi}\sin\psi\hat{D}^{-1}\sin\psi\ .
\ee
Let us further consider a complete set of normalized functions on
this interval (eigenfunctions of $\partial_\psi^2$),
\bea
\label{ea1.12}
f_n(\psi)=\left({2\over \pi-2\eta^{1/2}}\right)^{1/2}\sin
{n(\psi-\eta^{1/2})\over 1-(2/\pi)\eta^{1/2}}\ ;&& \\
\quad n=1,2,\ldots\ .&& \nonumber
\eea
It is easy to see that to the first order in $\eta^{1/2}$ we can
insert the projector $|f_1\rangle\langle f_1|$ in Eq.~(\ref{ea1.11}),
\be
\label{ea1.13}
\sigma_{xx}=e^2\nu v_F^2\tau{1\over\pi}
\langle\sin\psi|f_1\rangle^2\langle f_1|\hat{D}^{-1}|f_1\rangle
\ ,
\ee
where $$\langle f|g\rangle = \int_{\eta^{1/2}}^{\pi-\eta^{1/2}}d\psi
f(\psi)g(\psi)\ .$$
With the same accuracy we have
\be
\label{ea1.14}
\langle f_1|\hat{D}^{-1}|f_1\rangle = \langle
f_1|\hat{D}|f_1\rangle^{-1}\ .
\ee
Expanding everything to the first order in $\eta^{1/2}$, we find 
\bea
&& 
\langle f_1|\hat{D}|f_1\rangle = 1+
{4\over\pi}(1-C_1)\eta^{1/2}+O(\eta)\ , \label{ea1.15}\\
&&\langle f_1|\sin\psi\rangle=\left({\pi\over 2}\right)^{1/2}+O(\eta)\ ,
\label{ea1.16}
\eea
where 
$$C_1=\int_1^\infty dx\int_0^{2\pi}{d\vartheta\over 2\pi}
\left[1-\left(1-{\cos\vartheta\over x^2}\right)^{1/2}\right]\simeq
0.0241\ . $$
Substituting (\ref{ea1.14}), (\ref{ea1.15}), (\ref{ea1.16}) in 
(\ref{ea1.13}), we finally get
\be
\label{ea1.17}
\sigma_{xx}\simeq\sigma_0[1-C_2\eta^{1/2}]\ ,
\ee
with a numerical coefficient $C_2={4\over\pi}(1-C_1)\simeq
1.24$. Since at zero magnetic field $\rho_{xx}=\sigma_{xx}^{-1}$, we
arrive at the result (\ref{e2.38}) for the grating-induced correction
to resistivity.

\end{multicols}

\end{document}